\documentclass[conference]{IEEEtran}
\IEEEoverridecommandlockouts
\usepackage{cite}
\usepackage{amsmath,amssymb,amsfonts}
\usepackage{algorithmic}
\usepackage{textcomp}
\usepackage{xcolor}
\usepackage{orcidlink}
\usepackage{graphicx}
\usepackage{subcaption}
\usepackage{mathtools} 
\DeclarePairedDelimiter\ceil{\lceil}{\rceil}

\def\BibTeX{{\rm B\kern-.05em{\sc i\kern-.025em b}\kern-.08em
    T\kern-.1667em\lower.7ex\hbox{E}\kern-.125emX}}
\begin{document}

\title{Uncertainty-Aware Solar Flare Regression}

\author{\IEEEauthorblockN{Jinsu Hong\orcidlink{0009-0002-4383-1376}}
\IEEEauthorblockA{\textit{Computer Science} \\
\textit{Georgia State University}\\
Atlanta, USA \\
jhong36@gsu.edu
}
\and
\IEEEauthorblockN{Chetraj Pandey\orcidlink{0000-0002-4699-4050}}
\IEEEauthorblockA{\textit{Computer Science} \\
\textit{Texas Christian University}\\
Texas, USA \\
c.pandey@tcu.edu}
\and
\IEEEauthorblockN{Berkay Aydin\orcidlink{0000-0002-9799-9265}}
\IEEEauthorblockA{\textit{Computer Science} \\
\textit{Georgia State University}\\
Atlanta, USA \\
baydin2@gsu.edu}
}

\maketitle

\begin{abstract}
 Current solar flare predictions often lack precise quantification of their reliability, resulting in frequent false alarms, particularly when dealing with datasets skewed towards extreme events. To improve the trustworthiness of space weather forecasting, it is crucial to establish confidence intervals for model predictions. Conformal prediction, a machine learning framework, presents a promising avenue for this purpose by constructing prediction intervals that ensure valid coverage in finite samples without making assumptions about the underlying data distribution. In this study, we explore the application of conformal prediction to regression tasks in space weather forecasting. Specifically, we implement full-disk solar flare prediction using images created from magnetic field maps and adapt four pre-trained deep learning models to incorporate three distinct methods for constructing confidence intervals: conformal prediction, quantile regression, and conformalized quantile regression. Our experiments demonstrate that conformalized quantile regression achieves higher coverage rates and more favorable average interval lengths compared to alternative methods, underscoring its effectiveness in enhancing the reliability of solar weather forecasting models.
\end{abstract}


\section{Introduction}
A solar flare is a large eruption of electromagnetic radiation originating from the solar atmosphere. Large solar flares are often associated with coronal mass ejections (CMEs) \cite{Zhang2001}. Although Earth's atmosphere blocks most electromagnetic radiation from solar flares, high-energy radiation can pose a hazard to astronauts and instruments in space \cite{angryk2020multivariate}. Additionally, CMEs can cause severe power outages and disrupt communication systems by inducing currents in Earth's atmosphere \cite{Ahmed2011}. Therefore, accurate space weather forecasting is critical for protecting technological infrastructure and ensuring human safety.

Solar flares are categorized into five main classes (A, B, C, M, X in ascending order) with additional subdivisions denoted by floating-point values, such as M2.4 ($2.4 \times 10^{-5}\, W/m^{2}$) \cite{Fletcher2011}. Solar flares of M-class and above are of particular interest due to their significant impact on space weather and technological systems. However, these events are relatively rare, resulting in a highly imbalanced dataset \cite{abed2021automated}. Current solar flare prediction systems often suffer from high false alarm rates. To address this issue, various strategies have been explored, including sampling methods, data augmentation, custom loss functions \cite{pandey2024embedding}, and alternative labeling systems \cite{hong2023enhancing}. These existing solar flare prediction models provide point predictions, where the output is a single value. Although these models are often evaluated with global metrics that optimize overall performance and assess generalization, these metrics fail to provide confidence estimates for individual predictions. This lack of confidence assessment limits the ability to gauge the reliability of predictions, which is critical in high-stakes applications such as medical diagnostics, toxic drug discovery, and solar event forecasting. In these domains, providing a prediction interval with a confidence level for each instance is essential for informed decision-making. Specifically, prediction intervals not only provide valid coverage but also indicate how well a given instance is represented by the training space. For example, input data similar to the training set typically results in smaller intervals, while dissimilar data leads to larger intervals.

In this study, we focus on the application of conformal prediction, a machine learning framework that generates prediction intervals with user-defined confidence levels \cite{angelopoulos2021gentle}. Prediction intervals have the potential to reduce false alarms in solar event forecasting and support stakeholder decision-making. Specifically, we investigate the benefits of using conformal prediction and compare it to quantile regression \cite{koenker1978regression}. We also analyze the variation in prediction interval sizes across models with differing levels of complexity. Our findings demonstrate that conformalized quantile regression \cite{CQR2019} produces smaller prediction intervals while ensuring valid minimum coverage. The key contributions of this study are as follows: (i) Introducing a novel application of conformal prediction for solar flare regression. (ii) Evaluating the performance of various deep learning models for solar flare prediction and identifying the most effective model. (iii) Proposing a new evaluation method for assessing the performance of confidence intervals in predictive models. By systematically exploring the optimal combination of deep learning models and uncertainty quantification methods, this study aims to advance solar flare forecasting and enhance understanding of uncertainty in predictive modeling. 

The paper is structured as follows: Section \hyperref[related]{2} reviews related work and provides background on conformal prediction. Section \hyperref[data-prepar]{3} describes the data preparation process in detail. Section \hyperref[model]{4} outlines the architectures of the deep learning models explored in this study, along with their variations for conformal prediction and quantile regression. Section \hyperref[Experiments]{5} outlines the experimental design and introduces a novel evaluation method for conformal prediction. Finally, Section \hyperref[conclusion]{6} summarizes our findings and discusses future research directions.

\section{Background and Related work} \label{related}
\subsection{Full-disk Solar Flare Prediction}
Solar flares, as magnetic events, primarily originate from active regions (ARs) with strong and sometimes distorted magnetic fields. Specifically, they emerge from magnetic fields near sunspots within these active regions \cite{Li2020}. Magnetic observational data is commonly used to predict solar flares.

A full-disk solar flare model provides predictions for the entire solar disk, considering all active regions in the rasters, regardless of their direct association with solar flares \cite{pandey2021solar}. Full-disk images or rasters—whether point-in-time snapshots or time series—are linked to a defined prediction window, typically spanning 12, 24, or 48 hours, to forecast significant flare events. In this study, we focus on full-disk flare prediction models leveraging line-of-sight magnetograms with a 24-hour prediction window.

Another common approach to solar flare prediction is active region-based modeling, where each active region’s data is used to forecast solar events. These models represent active regions as point-in-time vectors \cite{ahmadzadeh2019challenges}, time series \cite{ji2022solar}, or images \cite{abed2021automated}. Time series classification techniques are often employed, and flares are labeled based on their association with specific active regions \cite{ji2020all}. Hybrid approaches combining full-disk and active region-based methods also exist in the literature \cite{pandey2022towards}.

While above methodologies have shown promise, existing solar flare forecasting primarily concentrates on point prediction without adequately quantifying uncertainty. This oversight hampers the reliability and interpretability of predictions, hindering effective risk assessment and decision-making processes. In response, our research aims to bridge this critical gap by integrating conformal prediction techniques into a deep learning framework.

\subsection{Conformal Prediction}
Conformal prediction (CP) \cite{Vovk1999} \cite{Vovk2005} is a framework for constructing prediction regions for individual predictions. Given a confidence level (e.g., 90\% or 95\%), CP ensures that the prediction regions—intervals for regression tasks and label sets for classification tasks—contain the true labels with the specified probability. The size of these prediction regions provides intuitive interpretability: if a new instance closely resembles the training data, the prediction region tends to be smaller, indicating higher confidence in the prediction. Conversely, when a new instance differs significantly from the training data, the prediction region expands, reflecting greater uncertainty.

CP operates under the assumption of exchangeability, meaning that data instances are drawn from the same distribution and that their order does not affect the conformalization process. This assumption is less restrictive than the independent and identically distributed (i.i.d.) requirement, making CP applicable to a wide range of machine learning models.

There are two main types of conformal prediction: transductive conformal prediction (TCP) and inductive conformal prediction (ICP) \cite{toccaceli2019combination}. The key difference lies in the use of a calibration set and the number of computations required. In TCP, the model must be retrained for each test instance, making it computationally expensive, and it is not applicable for regression tasks \cite{alvarsson2021predicting}. In contrast, ICP introduces a separate calibration set, which is a portion of the training data, but requires training the model only once. Due to its lower computational cost and compatibility with regression tasks, ICP is used in this study.

In ICP, a machine learning model ($\hat{\mu}$) is first trained using feature vectors ($X_i$) and target values ($Y_i$) from the training set $I_1$ with a regression algorithm ($A$), as defined in Eq. \hyperref[cp_model]{1}:
\begin{equation}\label{cp_model}
    \hat{\mu} \leftarrow A(\{X_i, Y_i\} : i \in I_1)
\end{equation}
To quantify uncertainty, a nonconformity measure ($R_i$) is computed in the calibration set $I_2$:
\begin{equation}\label{cp_score}
    R_i = |Y_i - \hat{\mu}(X_i)|, \; i \in I_2
\end{equation}
In this work, we utilize the absolute residual ($R_i$), although alternative nonconformity measures \cite{Papadopoulos2011} can be utilized for uncertainty estimation. Given a significance level $\alpha$ (i.e., confidence level $1-\alpha$), we compute the empirical quantile $Q$:
\begin{equation}\label{cp_quantile}
    \begin{aligned}
        & Q_{1-\alpha}(R, I_2) \coloneq \\
        & \ceil{(1-\alpha)(1+1/|I_2|)} \text{-th quantile of }\{R_i : i \in I_2\}
    \end{aligned}
\end{equation}
where $n = |I_2|$ and $\ceil{\cdot}$ denotes the ceiling function. The calibration set size influences the quantile value, with larger sizes typically leading to more stable coverage rates. Finally, the prediction interval for a new instance ($X_{n+1}$) is:
\begin{equation}\label{cp_interval}
    \begin{aligned}
        C(X_{n+1}) = [\hat{\mu}(X_{n+1})-Q_{1-\alpha}(R, I_2), \\
        \;\hat{\mu}(X_{n+1})+Q_{1-\alpha}(R, I_2)]
    \end{aligned}
\end{equation}
This confidence interval ensures that our prediction interval $C(X_{n+1})$ has at least a $100\times(1-\alpha)\%$ probability of covering the target value ($Y_{n+1}$) in Eq. \hyperref[cov]{5}:
\begin{equation}\label{cov}
    \mathbb{P}\{Y_{n+1} \in C(X_{n+1})\} \geq 1 - \alpha
\end{equation}
However, it is important to acknowledge that conformal prediction has limitations, such as fixed interval size ($2 \times Q$), which means the interval length remains constant for all instances, irrespective of input value ($X_{n+1}$). Consequently, this approach may struggle to accurately describe heteroscedastic data.


\subsection{Quantile regression} \label{QR}
Quantile regression (QR) enables the construction of confidence intervals using quantile loss \cite{Koenker1978} (as shown in Eq. \hyperref[quantile_loss]{6}).
\begin{equation}
\label{quantile_loss}
L_{\alpha}(y, \hat{y}) = 
\left\{
    \begin{array}{lr}
        \alpha\times(y - \hat{y}), & \text{if}\;\; \hat{y} \leq y\\
        (1-\alpha)\times(\hat{y} - y), & \text{if}\;\; \hat{y} > y
    \end{array}
\right\}
\end{equation} 
This loss function assigns different weights to residuals based on the chosen quantile level $\alpha$. When $\alpha = 0.5$, it reduces to the mean absolute error. For $\alpha$ values below 0.5, the model penalizes overestimated predictions more heavily, enabling the estimation of asymmetric prediction intervals.

In traditional quantile regression, two separate models ($\hat{\mu}_{\alpha/2}, \hat{\mu}_{1-\alpha/2}$) are trained with different quantile levels to construct prediction intervals with a confidence level of ($1-\alpha$), as described in Eq. \hyperref[qr_model]{7}.
\begin{equation}
\label{qr_model} 
\{\hat{\mu}_{\alpha/2},\; \hat{\mu}_{1-\alpha/2}\} \leftarrow A({X_i, Y_i} : i \in I_1) \end{equation}
However, in deep learning, a single model can be used by configuring the output layer to have two neurons, each corresponding to a different quantile level \cite{Taylor2000}. For example, if the significance level is set to 0.1 (corresponding to a 90\% confidence level), the two neurons ($\hat{a}_{lower}, \hat{a}_{upper})$ estimate the 5th and 95th percentiles, resulting in the interval [$\hat{a}_{0.05}$, $\hat{a}_{0.95}$] (see Eq. \hyperref[qr_interval]{8}).
\begin{equation}\label{qr_interval} C(X_{n+1}) = [\hat{a}_{\alpha/2},\; \hat{a}_{1-\alpha/2}] \end{equation}
This approach allows confidence intervals to be obtained without multiple model training. Additionally, since the interval width adapts to input data variability, quantile regression can effectively model heteroskedasticity. However, despite its flexibility, since the prediction intervals are derived from model estimates, they do not inherently guarantee minimum valid coverage ($\geq 1-\alpha$).

\subsection{Conformalized quantile regression}
Conformalized Quantile Regression (CQR) \cite{CQR2019} integrates conformal prediction with quantile regression to construct more reliable prediction intervals. Like conformal prediction, CQR requires a calibration set ($I_2$) to assess model uncertainty. Initially, two separate models, $\hat{\mu}_{\alpha/2}$ and $\hat{\mu}_{1-\alpha/2}$, are trained using quantile loss, as defined in Sec. \hyperref[QR]{2.3}, each corresponding to a specific significance level ($\alpha$). Next, the conformity score ($R_i$) is computed to quantify the combined uncertainty of both models. As defined in Eq. \hyperref[cqr_score]{9}, the conformity score measures the distance between the ground truth and the nearest predicted boundary:
\begin{equation}\label{cqr_score} 
    R_i \coloneq \max\{\hat{\mu}_{\alpha/2} - Y_i,\; Y_i-\hat{\mu}_{1-\alpha/2}\} 
\end{equation}
Similar to conformal prediction, the ($1-\alpha$)-th quantile of the nonconformity scores in the calibration set ($I_2$) is then determined using Eq. \hyperref[cp_quantile]{3}:
Finally, the prediction intervals obtained from the two quantile models are adjusted using $Q_{1-\alpha}(R, I_2)$, which accounts for the empirical distribution of nonconformity scores:
\begin{equation}\label{cqr_interval} 
    \begin{aligned} 
    C(X_{n+1}) = [\hat{\mu}_{\alpha/2}(X_{n+1})-Q_{1-\alpha}(R, I_2), \\
    \;\hat{\mu}_{1-\alpha/2}(X_{n+1})+Q_{1-\alpha}(R, I_2)] 
    \end{aligned} 
\end{equation}
The value of $Q_{1-\alpha}(R, I_2)$ determines the necessary adjustment. If $Q_{1-\alpha}(R, I_2)$ is positive, it indicates that more than half of the ground truths in the calibration set fall outside the initial prediction intervals, necessitating an expansion of the interval. Conversely, if $Q_{1-\alpha}(R, I_2)$ is negative, it suggests that the intervals are overly conservative and should be narrowed. Through this refinement, CQR ensures that the resulting prediction intervals achieve the desired confidence level ($1-\alpha$), addressing the limitations of quantile regression.

\begin{figure}[t]
    \centering
    \includegraphics[width=0.95\linewidth]{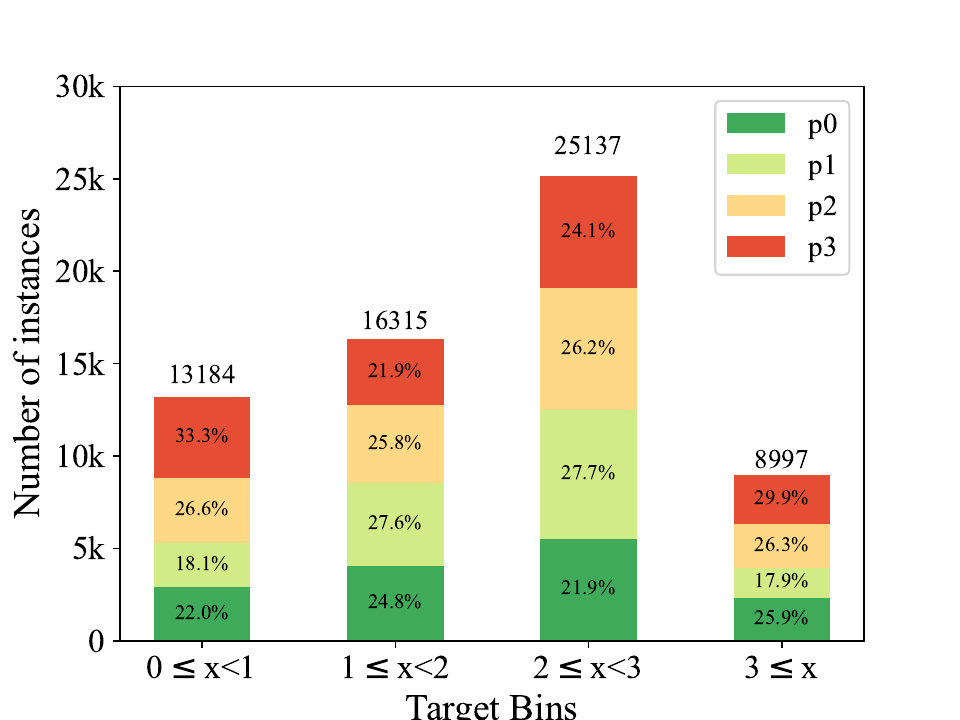}
    \caption{Distribution of normalized peak X-ray flux values (target values) across different partition.}
    \label{dist-of-label}
\end{figure}

\begin{figure*}[h!]
    \centering \label{CP-overview}
    \includegraphics[width=1\linewidth]{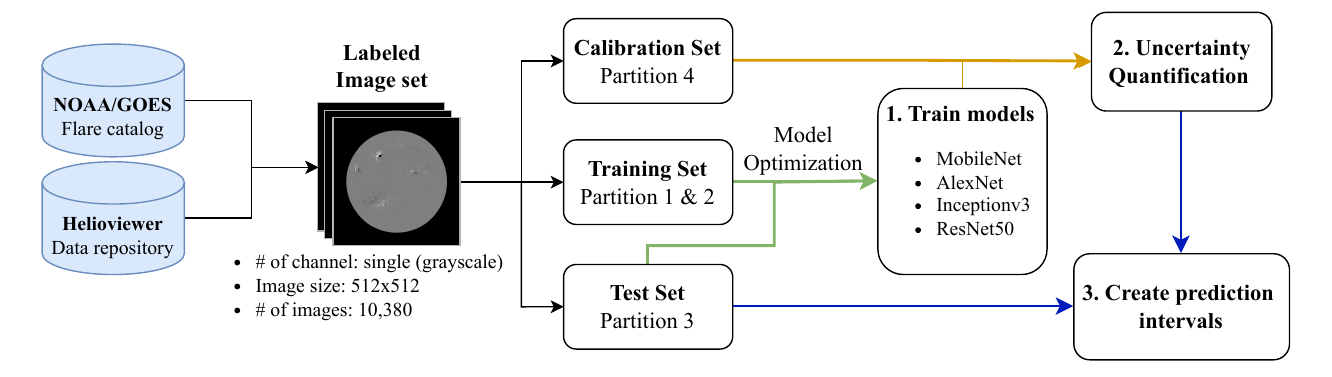}
    \caption{Overview of the process for conformal prediction.}
    \label{cv}
\end{figure*}

\section{Data Preparation}\label{data-prepar}
For full-disk solar flare prediction, we use Line-of-sight magnetogram images, sourced from the Helioseismic and Magnetic Imager (HMI) instrument onboard the Solar Dynamics Observatory (SDO). We retrieve the compressed versions (4096×4096 pixels) of these magnetograms from Helioviewer's public API \cite{helioviwer}. However, we resize the images to 512×512 pixels to reduce the feature dimensions and our images are not the original full-depth magnetic field rasters of range ±4500 Gauss but are compressed JP2 images (i.e., pixel values ranging from 0-255). These grayscale images, each with dimensions of 512x512 pixels, were captured at six-hour intervals spanning from December 6, 2010, to December 30, 2018, covering Solar Cycle 24 \cite{pandey2021solar}. The detailed code for generating 512×512 images is available in an anonymous repository.\footnote{Available at: \url{https://github.com/JinsuHongg/flare_reg_cp}}. We select four images from magnetograms at 00:00 UT, 06:00 UT 12:00 UT, and 18:00 UT.  For solar flare records, we use SunPy modules \cite{sunpy} to query data from the Heliophysics Events Knowledgebase (HEK) \cite{HEK}. To label these images accurately, we employ a sliding window approach. Each image is associated with the maximum intensity solar flare (and its peak X-ray flux) within the subsequent 24-hour window. We scan the records of the solar flares and find if there are solar flares within 24-hour window. When there are multiple solar flares in the 24-hour window, the label for the maximum intensity solar flare is chosen. Our dataset comprises total 10,380 data instances. 

We apply a logarithmic transformation to the X-ray flux values using the formula $y'_i = log_{10}(y_i) + 8$ to ensure that the intensity values are appropriately scaled. This adjustment ensures that intensity values fall within a range of [0, $\infty$]. Flare-quiet instances are labeled as 0, A-class flares are labeled between (0, 1), B- between [1, 2), C- between [2, 3), M- between [3, 4), X- between [4, $\infty$]. The full-disk solar flare images are divided into four subsets, each corresponding to a three-month period: Partition 1 (January–March), Partition 2 (April–June), Partition 3 (July–September), and Partition 4 (October–December) \cite{pandey2021solar}. We use this partitioning strategy because our images are time-dependent, with each image being similar to the one captured one hour later. Additionally, solar events, such as solar flares, exhibit cyclical patterns. By partitioning the data in this manner rather than randomly, we ensure a more evenly distributed dataset that better accounts for these temporal dependencies. The detailed procedure for creating labels for flare forecasting can be found in an anonymous repository.\footnote{Available at \url{https://github.com/JinsuHongg/flare_reg_cp/blob/main/training/data_split/labelmaker.py}} The distribution of target values across these partitions is depicted in Figure \hyperref[dist-of-label]{1}. To visualize the data distribution, the bins are defined as $[0,1)$, $[1,2)$, $[2,3)$, and $[3,\infty)$, corresponding to GOES classes A, B, C, and M or above, respectively. 

To ensure robust model training and evaluation, we split the dataset into three subsets: training, calibration, and testing set. The training set forms the basis for training the regression models, allowing it to discern underlying patterns in the magnetogram images and their corresponding solar flare intensities. The calibration set is crucial for assessing the model's uncertainty, measuring absolute residuals between predictions and actual labels. Lastly, the testing set serves as an independent evaluation dataset, confirming the model's ability to accurately predict solar flare occurrences on unseen data. Here, we utilize Partitions 1 and 2 for training, Partition 3 for testing, and Partition 4 for calibration as depicted in Figure \hyperref[CP-overview]{2}.

\section{Model Architecture}\label{model}
The task of predicting solar flares falls within the domain of supervised learning, given the availability of labeled data. For this study, we explore the efficacy of four deep learning algorithms: AlexNet, InceptionV3, MobileNet, and ResNet50. AlexNet \cite{alexnet} is renowned for its deep convolutional architecture, comprising eight layers, including five convolutional layers and three fully connected layers. InceptionV3 \cite{inceptionv3} is characterized by its inception module architecture, facilitating efficient use of computational resources by employing multiple convolutional operations within each module. MobileNet \cite{mobilenet-v3} is designed to be lightweight and efficient, making it suitable for deployment on resource-constrained devices. ResNet50 \cite{resnet50}, a variant of the ResNet architecture, is distinguished by its residual connections, which mitigate the vanishing gradient problem and enable training of significantly deeper networks. We select these four deep learning models to investigate the influence of model complexity and architecture on solar flare prediction and the construction of prediction intervals. We anticipate that models with more complex structures may yield narrower confidence intervals and higher coverage rates due to their ability to capture intricate patterns in the data.

To adapt these models to our task, we introduce specific modifications to their architectures. Firstly, considering that our input images are grayscale, we incorporate a convolutional layer with a kernel size of 3 and a stride of 2 to convert them into three-channel images, ensuring compatibility with the pre-trained models. Additionally, we adjust the output size of the last fully connected layer in each model to three, aligning with our objective of constructing prediction intervals. Each neuron in the output layer is associated with a different quantile loss function at a distinct significance level: $1-\alpha/2$, 0.5, and $\alpha/2$, where $\alpha \in [0, 1)$. The total loss is calculated as the average of three different quantile losses. The first and last neurons are designed for quantile regression and conformalized quantile regression, respectively. These two neurons construct the upper ($1-\alpha/2$) and lower bounds ($\alpha/2$) of the prediction interval. The intervals ([$\hat{a}_{1-\alpha/2}$, $\hat{a}_{\alpha/2}$]) constructed by quantile regression (QR) are adjusted using conformalized quantile regression (CQR) as described in Eq. \hyperref[cqr_interval]{10}. In CQR, the correction value ($Q$) is determined by the $\frac{n+1}{n} \times (1-\alpha)$\textsuperscript{th} quantile of the nonconformity scores in the calibration set, which is applied to refine the prediction intervals. The second (middle) neuron generates point predictions by outputting the median of predicted values, providing a central estimate based on the given input data. Additionally, this neuron contributes to constructing conformal prediction (CP) intervals. Model uncertainty is quantified using absolute residual in Eq. \hyperref[cp_score]{2} and the empirical quantile $Q$ is also derived as descibed in Eq. \hyperref[cp_quantile]{3}. Additionally, to examine the effect of significance level ($\alpha$), we explore four different levels: 5\%, 10\%, 15\%, and 20\%. 

{\small\tabcolsep=8pt  
\begin{table*}[t!]
\centering
\caption{Optimal parameters for each model with each confidence level}
\label{hyper-parameters}
\begin{tabular}{ccccccc}
\hline
                       &                                                            &                  & \multicolumn{4}{c}{\textbf{Optimal Parameters}}                                                                             \\
                       & Search Space                                               & Confidence Level & AlexNet                        & MobileNet                    & InceptionV3                  & ResNet50                     \\ \hline
\textbf{Learning Rate} & $10^{-7}$ to $10^{-5}$ & 95\%             & $2.5\times10^{-7}$ & $10^{-6}$   & $5\times10^{-7}$ & $5\times10^{-5}$ \\
                       &                                                            & 90\%             & $5\times10^{-6}$   & $5\times10^{-6}$ & $10^{-6}$   & $10^{-5}$   \\
                       &                                                            & 85\%             & $10^{-6}$     & $6\times10^{-8}$ & $5\times10^{-7}$ & $10^{-5}$   \\
                       &                                                            & 80\%             & $5\times10^{-6}$   & $10^{-6}$   & $10^{-6}$   & $10^{-5}$   \\
\textbf{Weight Decay}  & 0, $10^{-5}$, $10^{-4}$  & 95\%             & $10^{-5}$     & $10^{-4}$   & 0                            & $10^{-4}$   \\
                       &                                                            & 90\%             & $10^{-5}$     & $10^{-5}$   & 0                            & $10^{-5}$   \\
                       &                                                            & 85\%             & $10^{-5}$     & $10^{-5}$   & 0                            & 0                            \\
                       &                                                            & 80\%             & $10^{-5}$     & $10^{-5}$   & 0                            & 0                            \\ \hline
\end{tabular}
\end{table*}
}
 
\section{EXPERIMENTS} \label{Experiments}
\subsection{Experimental Settings}
We employed four distinct deep learning architectures, including AlexNet, MobileNet, InceptionV3, and ResNet50, leveraging the 'Adam' optimizer \cite{kingma2014adam} for efficient parameter updates. To ensure robust convergence during training, we incorporated the 'ReduceLROnPlateau' learning rate scheduler from the PyTorch package. This scheduler dynamically adjusts learning rates when no performance improvement is detected over a predefined number of epochs. In our experiments, the initial learning rate is reduced by a factor of 0.5 if no improvement in loss is observed over 5 epochs. The parameters in this study are shown in Table \hyperref[hyper-parameters]{1}. Different learning rates and weight decays are used for different confidence level, because our total loss is vary depending on confidence levels ($1-\alpha$). To determine optimal learning rates and weight decay parameters, we evaluate R-squared metric (or coefficient of determination) between predicted outcomes and ground truth values.

In our evaluation process, we introduce a novel metric that simultaneously accounts for both interval length and marginal coverage rate. The interval length represents the mean prediction interval length in the test set, while the marginal coverage rate denotes the proportion of instances in the test set covered by the prediction interval. The proposed evaluation metric is a weighted sum of the normalized interval length and the marginal coverage rate. It assesses how small the interval length is relative to the range of target values and how closely the marginal coverage aligns with the confidence level. The ideal scenario occurs when the marginal coverage exactly matches the confidence level while keeping the average prediction interval as small as possible. As detailed in Eq. \hyperref[data-interval]{11} - Eq. \hyperref[ccindex]{15}, our evaluation metric begins by determining $L_{test}$ in Eq. \hyperref[data-interval]{11}, which represents range of target values ($Y_{test}$) in the test set ($I_{test}$). Similarly, $L_{Avg.}$ (Eq. \hyperref[L-avg]{12}) defines the average length of prediction intervals ($C(X_{test})$) in the test set. 
\begin{equation}\label{data-interval}
    L_{test} = Max(Y_{test}) - Min(Y_{test})
\end{equation}
\begin{equation} \label{L-avg}
    L_{Avg.} = \frac{1}{|I_{test}|}\sum|C(X_{i})|, \; i\in I_{test}
\end{equation}
We then calculate the $\gamma$ value in Eq. \hyperref[gamma]{13}, which signifies the ratio between the average prediction interval length ($L_{Avg.}$) and size of target feature space $L_{test}$, in this study $L_{test}$ is approximately 4.97. 
\begin{equation} \label{gamma}
    \gamma = \frac{L_{Avg.}}{L_{test}}
\end{equation}
The $\gamma$ value provides insight into how our average interval length compares to the size of the target space. A smaller $\gamma$ indicates better performance when the coverage rate is maintained. Additionally, we introduce the parameter $\delta$ to characterize the model's marginal coverage relative to the significance level ($\alpha$). 
\begin{equation}
    \delta = \frac{|\eta-(1-\alpha)|}{\alpha}
\end{equation}
Here, eta ($\eta$) represents the marginal coverage rate of the model in the test set, while $\alpha$ takes values of 0.05, 0.1, 0.15, and 0.2, corresponding to the chosen significance levels. We introduce the Confidence Coverage Index ($I_{CC}$), defined as one minus the weighted sum of the normalized interval length ($\gamma$) and the relative deviation in coverage ($\delta$). In our study, we assign equal weights to both components. The Confidence Coverage Index ($I_{CC}$) ranges within ( $-\infty$, 1].
\begin{equation}\label{ccindex}
    I_{CC} = 1 - (w*\delta + (1-w)*\gamma)
\end{equation}
Adjusting the weights ($w$) allows us to account for the relative importance of $\gamma$ and $\delta$ in determining the index, thus providing flexibility in its interpretation and application. The source code and experimental results for reproducing this work are available in our open-source repository (anonymous) 

\begin{table}[t!]
\centering \label{result-single}
\caption{The R-squared results of the models, with each learning algorithm trained using four different significance levels ($\alpha$). Among the models, ResNet50 demonstrates the highest average R-squared score.}
\begin{tabular}{lcc}
\hline
Model       & Average & Standard Deviation \\
\hline
AlexNet     & 0.473   & 0.040              \\
MobileNet   & 0.454   & 0.037              \\
InceptionV3 & 0.458   & 0.028              \\
ResNet50    & 0.496   & 0.017     \\ \hline        
\end{tabular}
\end{table}

\begin{table}[h!]
\centering \label{result-CP}
\caption{The results of average prediction lengths, marginal coverages, and $I_{CC}$ across different models and uncertainty quantification methods are presented. In this table, we report only the results for models with a significance level of 0.1.}
\begin{tabular}{ccccc}
\hline
    &             & \textbf{\begin{tabular}[c]{@{}c@{}}Avg. \\ Length\end{tabular}} & \textbf{\begin{tabular}[c]{@{}c@{}}Avg. \\ Coverage\end{tabular}} & \textbf{$I_{CC}$} \\
    \hline
CP  & \textbf{AlexNet}     & 2.39                                                                & 87.3                                                                       & \textbf{0.530}        \\
    & MobileNet   & 2.35                                                                & 86.9                                                                       & 0.516        \\
    & InceptionV3 & 2.49                                                                & 84.6                                                                       & 0.380        \\
    & ResNet50    & 2.33                                                                & 86.1                                                                       & 0.478        \\
QR  & AlexNet     & 2.38                                                                & 85.8                                                                       & 0.454        \\
    & MobileNet   & 2.08                                                                & 82.6                                                                       & 0.336        \\
    & InceptionV3 & 2.40                                                                & 74.9                                                                       & -0.091       \\
    & ResNet50    & 2.40                                                                & 79.6                                                                       & 0.142        \\
CQR & AlexNet     & 2.43                                                                & 86.4                                                                       & 0.476        \\
    & MobileNet   & 2.19                                                                & 84.6                                                                       & 0.423        \\
    & InceptionV3 & 2.96                                                                & 87.5                                                                       & 0.457        \\
    & ResNet50    & 2.73                                                                & 86.1                                                                       & 0.422       \\ \hline
\end{tabular}
\end{table}

\begin{figure*}[t!]
    \centering
    \includegraphics[width=0.90\linewidth]{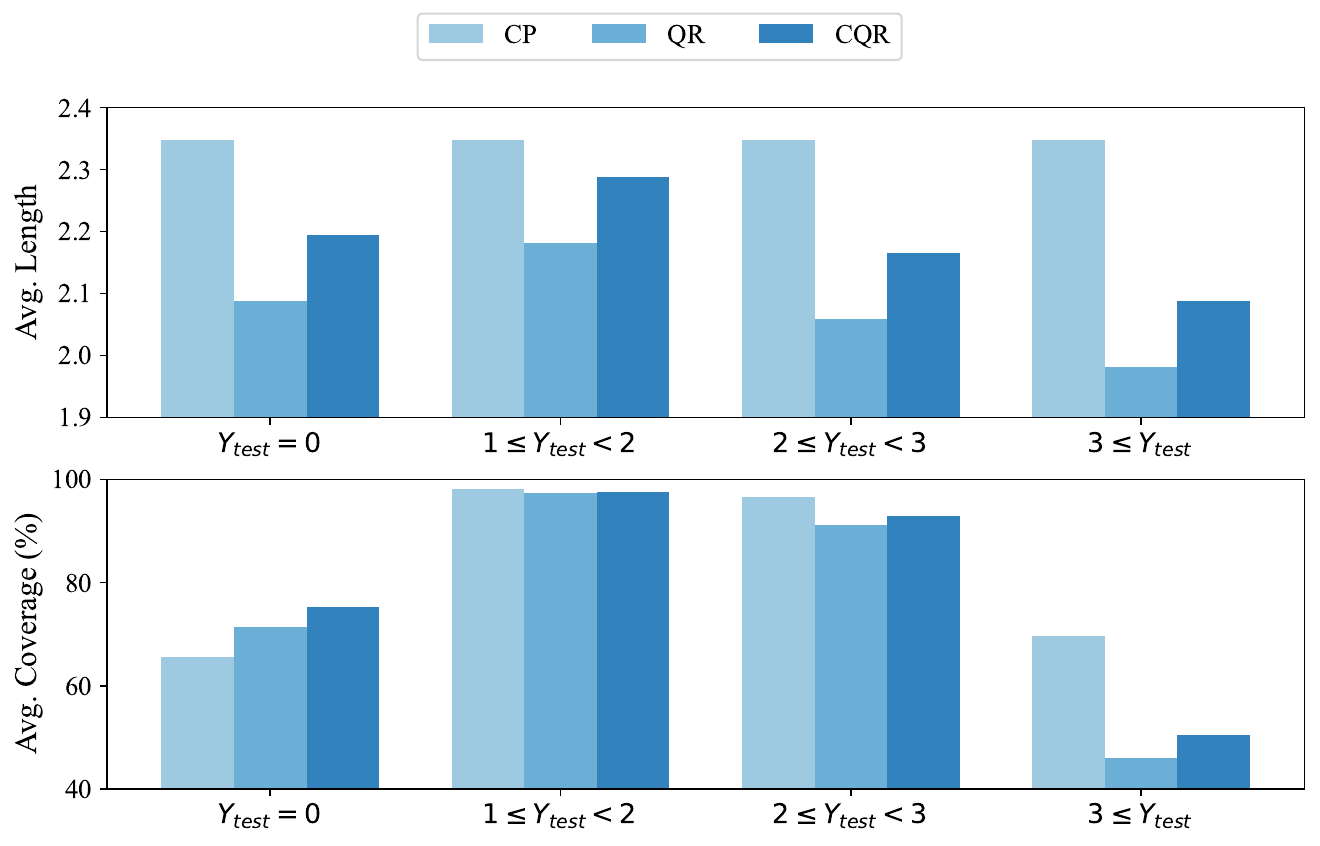}
    \caption{This figure presents the evaluation of prediction intervals from MobileNet. In this result, a 90\% confidence level and four different target bins are used to analyze three uncertainty quantification methods: CP, QR, and CQR. While QR and CQR produce narrower prediction intervals, they exhibit lower marginal coverage compared to CP. However, the prediction interval lengths from CP remain constant across all bins.}
    \label{MobileNet result}
\end{figure*}

\subsection{Result and Discussion}\label{result}
As explained Section \hyperref[data-prepar]{3}-\hyperref[Experiments]{5}, we predict maximum solar X-ray flux with 24-hour window using full-disk magnetograms. Our data is split into three parts (training, calibration, and testing) for quantifying models' uncertainty. Also, four different significance levels (5\%, 10\%, 15\%, and 20\%) are explored, and this requires each deep learning model is trained four times with different significance level ($\alpha$). After training and calibrating the models, the models are evaluated via marginal coverage, average prediction interval, and the confidence coverage index ($I_{CC}$). 

We first evaluate the performance of individual results from four deep learning models, as shown in Table \hyperref[result-single]{2}. For each model, we compute the mean and standard deviation of R-squared scores across four confidence levels: 80\%, 85\%, 90\%, and 95\%. Among the models, ResNet50 demonstrates the best overall performance. The comprehensive results of CP, QR, and CQR are tabulated from Table \hyperref[result-CP]{3}, encompassing the averaged interval length, coverage rate, and $I_{CC}$ for the models trained and calibrated with the confidence level of 0.9 (90\%). Unlike the point prediction results reported in Table \hyperref[result-single]{2}, our findings in Table \hyperref[result-CP]{3} suggest that models with lower computational complexity (AlexNet and MobileNet) outperform more complex architectures (InceptionV3 and ResNet50). Interestingly, our findings contradict our initial expectations since these simpler models—requiring fewer computational resources—outperformed more complex architectures. This discrepancy may be due to a limited number of instances or insufficient information in the magnetograms.

Additionally, in Table \hyperref[result-CP]{3}, conformalized quantile regression (CQR) achieves higher marginal coverage than quantile regression (QR), demonstrating the effectiveness of CQR. However, all three uncertainty quantification methods in Table \hyperref[result-CP]{3}, fail to meet the minimum coverage requirement ($1-\alpha$). A possible explanation for this is that when splitting the dataset into three parts (training, calibration, and test sets), we account for time dependency rather than performing a random split. This is because we sample the input magnetogram data at specific time intervals (6-hour intervals in this study). As a result, a distributional discrepancy arises between the calibration and test sets, leading to deviations in coverage. 

\begin{figure}[t]
    \centering \label{length-year}
    \includegraphics[width=0.95\linewidth]{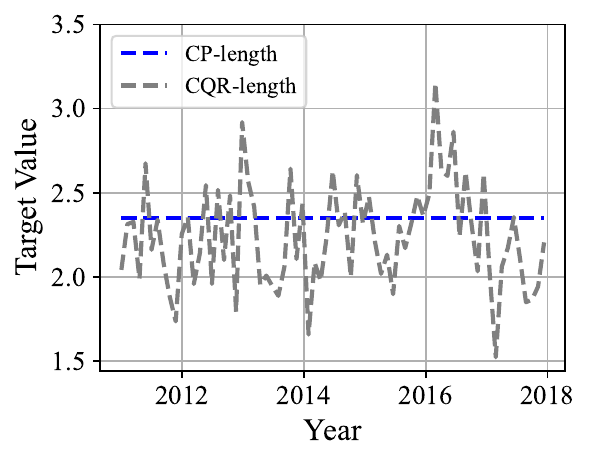}
    \caption{Comparison of prediction intervals generated by CP and CQR over an eight-year period.}
    \label{fig:enter-label}
\end{figure}

In terms of the $I_{CC}$ score, CP demonstrates the best performance. However, it's crucial to note that CP generates fixed interval lengths regardless of the input data, failing to account for the dataset's heteroskedasticity. Figure \hyperref[MobileNet result]{3} presents a detailed analysis of interval length and coverage rate. The bar plot specifically examines MobileNet with a 90\% confidence level to enable clear comparisons. We analyze CP, QR, and CQR across different target value bins ($Y_{test} = 0$, $0\leq Y_{test}<2$, $2\leq Y_{test}<3$, and $3 \leq Y_{test}$). In the upper bar plot of Figure \hyperref[MobileNet result]{3}, CP maintains consistent interval lengths, while QR and CQR show variation, demonstrating their capacity to capture data heteroskedasticity. Also, in Figure \hyperref[length-year]{4}, CP exhibits constant interval lengths regardless of the input data, while CQR demonstrates variable interval lengths that adapt to the underlying data distribution. Consequently, CP typically yields longer prediction intervals and higher coverage rates due to its uniform approach across all target values.

In the lower bar plot of Figure \hyperref[MobileNet result]{3}, we observe that the marginal coverage for two bins ($Y_{test}=0$ and $3\leq Y_{test}$) falls below 90\%. This occurs because both CP and CQR guarantee only marginal coverage rather than bin-specific coverage. While class-wise coverage in classification tasks can be achieved using Mondrian Conformal Prediction \cite{vovk2003mondrian}, ensuring valid coverage for each bin in regression tasks requires segmenting target values and quantifying uncertainties within individual bins \cite{bostrom2020mondrian}. This approach presents an interesting direction for future research.

Determining the optimal coverage rate presents a unique challenge, as our confidence coverage index ($I_{CC}$) tends to increase with higher significance levels. Ideally, our prediction intervals should have a length less than two, ensuring that they remain within adjacent solar flare classes. Our analysis suggests that a significance level of 20\% guarantees prediction intervals below this threshold, striking a balance between coverage and precision in our predictions.

\section{Conclusion and Future work} \label{conclusion}
In this study, we conducted a comprehensive evaluation of four deep learning models across different significance levels and uncertainty quantification methods. Notably, AlexNet emerged as the top performer, achieving the highest confidence coverage index score among all models. However, our findings challenge conventional expectations, as simpler models, such as MobileNet and AlexNet, consistently outperformed more complex architectures, as shown in Table \hyperref[result-CP]{3}. We suspect that we need a larger dataset by incorporating additional channel images or extending the dataset range from 2010–2018 to 2010–2025. Alternatively, we could explore different target values instead of peak X-ray flux.

When comparing methods for constructing prediction intervals, we found that conformal prediction (CP) demonstrated robust performance, as evidenced by Table \hyperref[result-CP]{3}. However, it struggled to capture the heteroskedasticity in our dataset, highlighting areas for improvement. In contrast, conformalized quantile regression (CQR) achieved comparable results to CP but exhibited slightly lower marginal coverage rates. In Figure \hyperref[MobileNet result]{3}, the coverage rate drops when the target value exceeds three. This occurs because instances with target values greater than three do not contribute sufficiently to the computation of score functions in the calibration set, making it difficult to determine the proper $(1-\alpha)$th quantile of the scores. Addressing this imbalance requires implementing sampling techniques, particularly in the calibration set, and exploring different methods for computing nonconformity scores. For future work, we plan to discretize the target values into several bins and compute nonconformity scores based on these bins. Additionally, we will explore various sampling methods to enhance coverage rates for underrepresented labels.

\begin{table}[]
\centering
\label{cov-len}
\caption{The result of MobileNet with four different confidence levels.  }
\begin{tabular}{cccccc}
\hline
         & Confidence & 80\% & 85\% & 90\% & 95\% \\ \hline
         & CP         & 1.78 & 1.92 & 2.35 & 2.98 \\
Length   & QR         & 1.36 & 1.83 & 2.08 & 2.34 \\
         & CQR        & 1.60 & 1.87 & 2.19 & 2.70 \\
         & CP         & 76.5 & 78.6 & 86.9 & 90.9 \\
Coverage & QR         & 67.1 & 75.9 & 82.6 & 85.4 \\
         & CQR        & 74.3 & 77.2 & 84.6 & 89.9 \\
         & CP         & 0.14 & 0.46 & 0.54 & 0.67 \\
$I_{CC}$ & QR         & 0.27 & 0.47 & 0.52 & 0.52 \\
         & CQR        & 0.41 & 0.50 & 0.54 & 0.61 \\ \hline
\end{tabular}
\end{table}

\bibliographystyle{IEEEtran}
\bibliography{ref}

@article{Zhang2001,
  title = {On the Temporal Relationship between Coronal Mass Ejections and Flares},
  volume = {559},
  ISSN = {1538-4357},
  number = {1},
  journal = {The Astrophysical Journal},
  publisher = {American Astronomical Society},
  author = {Zhang,  J. and Dere,  K. P. and Howard,  R. A. and Kundu,  M. R. and White,  S. M.},
  year = {2001},
  month = sep,
  pages = {452–462}
}

@article{Ahmed2011,
  title = {Solar Flare Prediction Using Advanced Feature Extraction,  Machine Learning,  and Feature Selection},
  volume = {283},
  ISSN = {1573-093X},
  number = {1},
  journal = {Solar Physics},
  publisher = {Springer Science and Business Media LLC},
  author = {Ahmed,  Omar W. and Qahwaji,  Rami and Colak,  Tufan and Higgins,  Paul A. and Gallagher,  Peter T. and Bloomfield,  D. Shaun},
  year = {2011},
  month = nov,
  pages = {157–175}
}

@ARTICLE{helioviwer,
       author = {{Muller}, D. and {Fleck}, B. and {Dimitoglou}, G. and {Caplins}, B.~W. and {Amadigwe}, D.~E. and {Garc{\'\i}a Ortiz}, J.~P. and {Wamsler}, B. and {Alexanderian}, A. and {Hughitt}, V.~K. and {Ireland}, J.},
        title = "{JHelioviewer: Visualizing Large Sets of Solar Images Using JPEG 2000}",
      journal = {Computing in Science and Engineering},
         year = 2009,
        month = sep,
       volume = {11},
       number = {5},
        pages = {38-47},
archivePrefix = {arXiv},
       eprint = {0906.1582},
 primaryClass = {astro-ph.SR}
}

@article{Fletcher2011,
  title = {An Observational Overview of Solar Flares},
  volume = {159},
  ISSN = {1572-9672},
  url = {http://dx.doi.org/10.1007/s11214-010-9701-8},
  DOI = {10.1007/s11214-010-9701-8},
  number = {1–4},
  journal = {Space Science Reviews},
  publisher = {Springer Science and Business Media LLC},
  author = {Fletcher,  L. and Dennis,  B. R. and Hudson,  H. S. and Krucker,  S. and Phillips,  K. and Veronig,  A. and Battaglia,  M. and Bone,  L. and Caspi,  A. and Chen,  Q. and Gallagher,  P. and Grigis,  P. T. and Ji,  H. and Liu,  W. and Milligan,  R. O. and Temmer,  M.},
  year = {2011},
  month = aug,
  pages = {19–106}
}

@article{angryk2020multivariate,
  title={Multivariate time series dataset for space weather data analytics},
  author={Angryk, Rafal A and Martens, Petrus C and Aydin, Berkay and Kempton, Dustin and Mahajan, Sushant S and Basodi, Sunitha and Ahmadzadeh, Azim and Cai, Xumin and Filali Boubrahimi, Soukaina and Hamdi, Shah Muhammad and others},
  journal={Scientific data},
  volume={7},
  number={1},
  pages={227},
  year={2020},
  publisher={Nature Publishing Group UK London}
}

@inproceedings{pandey2024embedding,
  title={Embedding Ordinality to Binary Loss Function for Improving Solar Flare Forecasting},
  author={Pandey, Chetraj and Ji, Anli and Hong, Jinsu and Angryk, Rafal A and Aydin, Berkay},
  booktitle={2024 IEEE 11th International Conference on Data Science and Advanced Analytics (DSAA)},
  pages={1--10},
  year={2024},
  organization={IEEE}
}

@inproceedings{hong2023enhancing,
  title={Enhancing Solar Flare Prediction with Innovative Data-Driven Labels},
  author={Hong, Jinsu and Ji, Anli and Pandey, Chetraj and Aydin, Berkay},
  booktitle={2023 IEEE 5th International Conference on Cognitive Machine Intelligence (CogMI)},
  pages={190--195},
  year={2023},
  organization={IEEE}
}

@article{kingma2014adam,
  title={Adam: A method for stochastic optimization},
  author={Kingma, Diederik P},
  journal={arXiv preprint arXiv:1412.6980},
  year={2014}
}

@article{Li2020,
  title = {Predicting Solar Flares Using a Novel Deep Convolutional Neural Network},
  volume = {891},
  ISSN = {1538-4357},
  number = {1},
  journal = {The Astrophysical Journal},
  publisher = {American Astronomical Society},
  author = {Li,  Xuebao and Zheng,  Yanfang and Wang,  Xinshuo and Wang,  Lulu},
  year = {2020},
  month = feb,
  pages = {10}
}

@inproceedings{pandey2021solar,
  title={Solar Flare Forecasting with Deep Neural Networks using Compressed Full-disk HMI Magnetograms},
  author={Pandey, Chetraj and Angryk, Rafal A and Aydin, Berkay},
  booktitle={2021 IEEE International Conference on Big Data (Big Data)},
  pages={1725--1730},
  year={2021},
  organization={IEEE}
}

@inproceedings{ji2020all,
  title={All-clear flare prediction using interval-based time series classifiers},
  author={Ji, Anli and Aydin, Berkay and Georgoulis, Manolis K and Angryk, Rafal},
  booktitle={2020 IEEE International Conference on Big Data (Big Data)},
  pages={4218--4225},
  year={2020},
  organization={IEEE}
}

@inproceedings{ahmadzadeh2019challenges,
  title={Challenges with extreme class-imbalance and temporal coherence: A study on solar flare data},
  author={Ahmadzadeh, Azim and Hostetter, Maxwell and Aydin, Berkay and Georgoulis, Manolis K and Kempton, Dustin J and Mahajan, Sushant S and Angryk, Rafal},
  booktitle={2019 IEEE international conference on big data (Big Data)},
  pages={1423--1431},
  year={2019},
  organization={Ieee}
}

@inproceedings{ji2022solar,
  title={Solar Flare Forecasting with Deep Learning-based Time Series Classifiers},
  author={Ji, Anli and Wen, Junzhi and Angryk, Rafal and Aydin, Berkay},
  booktitle={2022 26th International Conference on Pattern Recognition (ICPR)},
  pages={2907--2913},
  year={2022},
  organization={IEEE}
}

@article{abed2021automated,
  title={The automated prediction of solar flares from SDO images using deep learning},
  author={Abed, Ali K and Qahwaji, Rami and Abed, Ahmed},
  journal={Advances in Space Research},
  volume={67},
  number={8},
  pages={2544--2557},
  year={2021},
  publisher={Elsevier}
}

@article{pandey2022towards,
  title={Towards coupling full-disk and active region-based flare prediction for operational space weather forecasting},
  author={Pandey, Chetraj and Ji, Anli and Angryk, Rafal A and Georgoulis, Manolis K and Aydin, Berkay},
  journal={Frontiers in Astronomy and Space Sciences},
  volume={9},
  pages={897301},
  year={2022},
  publisher={Frontiers}
}

@inproceedings{Vovk1999,
title = {Machine Learning Applications of Algorithmic Randomness},
author = {Alex Gammerman, Craig Saunders},
year = {1999},
isbn = {1558606122},
publisher = {Morgan Kaufmann Publishers Inc.},
address = {San Francisco, CA, USA},
booktitle = {Proceedings of the Sixteenth International Conference on Machine Learning},
pages = {444–453},
numpages = {10},
series = {ICML '99}
}

@book{Vovk2005,
author = {Vovk, Vladimir and Gammerman, Alex and Shafer, Glenn},
title = {Algorithmic Learning in a Random World},
year = {2005},
isbn = {0387001522},
publisher = {Springer-Verlag},
address = {Berlin, Heidelberg}
}

@article{Koenker1978,
  title = {Regression Quantiles},
  volume = {46},
  ISSN = {0012-9682},
  number = {1},
  journal = {Econometrica},
  publisher = {JSTOR},
  author = {Koenker,  Roger and Bassett,  Gilbert},
  year = {1978},
  month = jan,
  pages = {33}
}

@article{Taylor2000,
author = {Taylor, James W.},
title = {A quantile regression neural network approach to estimating the conditional density of multiperiod returns},
journal = {Journal of Forecasting},
volume = {19},
number = {4},
pages = {299-311},
year = {2000}
}

@inproceedings{CQR2019,
 author = {Romano, Yaniv and Patterson, Evan and Candes, Emmanuel},
 booktitle = {Advances in Neural Information Processing Systems},
 editor = {H. Wallach and H. Larochelle and A. Beygelzimer and F. d\textquotesingle Alch\'{e}-Buc and E. Fox and R. Garnett},
 pages = {},
 publisher = {Curran Associates, Inc.},
 title = {Conformalized Quantile Regression},
 volume = {32},
 year = {2019}
}

@article{Papadopoulos2011,
  title = {Regression Conformal Prediction with Nearest Neighbours},
  volume = {40},
  ISSN = {1076-9757},
  journal = {Journal of Artificial Intelligence Research},
  publisher = {AI Access Foundation},
  author = {Papadopoulos,  H. and Vovk,  V. and Gammerman,  A.},
  year = {2011},
  month = apr,
  pages = {815–840}
}

@article{alvarsson2021predicting,
  title={Predicting with confidence: using conformal prediction in drug discovery},
  author={Alvarsson, Jonathan and McShane, Staffan Arvidsson and Norinder, Ulf and Spjuth, Ola},
  journal={Journal of Pharmaceutical Sciences},
  volume={110},
  number={1},
  pages={42--49},
  year={2021},
  publisher={Elsevier}
}

@article{toccaceli2019combination,
  title={Combination of inductive mondrian conformal predictors},
  author={Toccaceli, Paolo and Gammerman, Alexander},
  journal={Machine Learning},
  volume={108},
  pages={489--510},
  year={2019},
  publisher={Springer}
}

@article{angelopoulos2021gentle,
  title={A gentle introduction to conformal prediction and distribution-free uncertainty quantification},
  author={Angelopoulos, Anastasios N and Bates, Stephen},
  journal={arXiv preprint arXiv:2107.07511},
  year={2021}
}

@article{koenker1978regression,
  title={Regression quantiles},
  author={Koenker, Roger and Bassett Jr, Gilbert},
  journal={Econometrica: journal of the Econometric Society},
  pages={33--50},
  year={1978},
  publisher={JSTOR}
}

@article{vovk2003mondrian,
  title={Mondrian confidence machine},
  author={Vovk, Vladimir and Lindsay, David and Nouretdinov, Ilia and Gammerman, Alex},
  journal={Technical Report},
  year={2003}
}

@inproceedings{bostrom2020mondrian,
  title={Mondrian conformal regressors},
  author={Bostr{\"o}m, Henrik and Johansson, Ulf},
  booktitle={Conformal and Probabilistic Prediction and Applications},
  pages={114--133},
  year={2020},
  organization={PMLR}
}

@article{mobilenet-v3,
  author       = {Andrew Howard and
                  Mark Sandler and
                  Grace Chu and
                  Liang{-}Chieh Chen and
                  Bo Chen and
                  Mingxing Tan and
                  Weijun Wang and
                  Yukun Zhu and
                  Ruoming Pang and
                  Vijay Vasudevan and
                  Quoc V. Le and
                  Hartwig Adam},
  title        = {Searching for MobileNetV3},
  journal      = {CoRR},
  volume       = {abs/1905.02244},
  year         = {2019},

  eprinttype    = {arXiv},
  eprint       = {1905.02244},
}

@article{alexnet,
  title = {ImageNet classification with deep convolutional neural networks},
  volume = {60},
  ISSN = {1557-7317},
  number = {6},
  journal = {Communications of the ACM},
  publisher = {Association for Computing Machinery (ACM)},
  author = {Krizhevsky,  Alex and Sutskever,  Ilya and Hinton,  Geoffrey E.},
  year = {2017},
  month = may,
  pages = {84–90}
}

@inproceedings{inceptionv3,
  title = {Rethinking the Inception Architecture for Computer Vision},
  booktitle = {2016 IEEE Conference on Computer Vision and Pattern Recognition (CVPR)},
  publisher = {IEEE},
  author = {Szegedy,  Christian and Vanhoucke,  Vincent and Ioffe,  Sergey and Shlens,  Jon and Wojna,  Zbigniew},
  year = {2016},
  month = jun 
}

@INPROCEEDINGS{resnet50,
  author={He, Kaiming and Zhang, Xiangyu and Ren, Shaoqing and Sun, Jian},
  booktitle={2016 IEEE Conference on Computer Vision and Pattern Recognition (CVPR)}, 
  title={Deep Residual Learning for Image Recognition}, 
  year={2016},
  volume={},
  number={},
  pages={770-778},
  }

@article{sunpy,
doi = {10.1088/1749-4699/8/1/014009},
url = {https://dx.doi.org/10.1088/1749-4699/8/1/014009},
year = {2015},
month = {jul},
publisher = {IOP Publishing},
volume = {8},
number = {1},
pages = {014009},
author = {The SunPy Community and Stuart J Mumford and Steven Christe and David Pérez-Suárez and Jack Ireland and Albert Y Shih and Andrew R Inglis and Simon Liedtke and Russell J Hewett and Florian Mayer and Keith Hughitt and Nabil Freij and Tomas Meszaros and Samuel M Bennett and Michael Malocha and John Evans and Ankit Agrawal and Andrew J Leonard and Thomas P Robitaille and Benjamin Mampaey and Jose Iván Campos-Rozo and Michael S Kirk},
title = {SunPy—Python for solar physics},
journal = {Computational Science \& Discovery},
}

@article{HEK,
  title = {Heliophysics Event Knowledgebase for the Solar Dynamics Observatory (SDO) and Beyond},
  volume = {275},
  ISSN = {1573-093X},
  url = {http://dx.doi.org/10.1007/s11207-010-9624-2},
  DOI = {10.1007/s11207-010-9624-2},
  number = {1–2},
  journal = {Solar Physics},
  publisher = {Springer Science and Business Media LLC},
  author = {Hurlburt,  N. and Cheung,  M. and Schrijver,  C. and Chang,  L. and Freeland,  S. and Green,  S. and Heck,  C. and Jaffey,  A. and Kobashi,  A. and Schiff,  D. and Serafin,  J. and Seguin,  R. and Slater,  G. and Somani,  A. and Timmons,  R.},
  year = {2010},
  month = dec,
  pages = {67–78}
}


\appendix
\section{Additional tables}

\begin{table*}[ht!] \label{CP-6h}
\caption{CP result}
\centering
\begin{tabular}{ccccccccc}
\textbf{Error rate} &
  \textbf{5\%} &
  \textbf{10\%} &
  \textbf{15\%} &
  \textbf{20\%} &
  \textbf{5\%} &
  \textbf{10\%} &
  \textbf{15\%} &
  \textbf{20\%} \\ \hline
\textbf{\begin{tabular}[c]{@{}c@{}}Target \\ values\end{tabular}} &
  \multicolumn{4}{c|}{$Y_{all}$} &
  \multicolumn{4}{c}{$Y_{target}\geq3$} \\ \hline
\textbf{Mobilenet} &
  \begin{tabular}[c]{@{}c@{}}2.98 \\ (90.9\%)\end{tabular} &
  \begin{tabular}[c]{@{}c@{}}2.35 \\ (86.9\%)\end{tabular} &
  \begin{tabular}[c]{@{}c@{}}1.92 \\ (78.6\%)\end{tabular} &
  \multicolumn{1}{c|}{\begin{tabular}[c]{@{}c@{}}1.78 \\ (76.5\%)\end{tabular}} &
  \begin{tabular}[c]{@{}c@{}}2.98 \\ (86.1\%)\end{tabular} &
  \begin{tabular}[c]{@{}c@{}}2.35 \\ (69.7\%)\end{tabular} &
  \begin{tabular}[c]{@{}c@{}}1.92 \\ (47.0\%)\end{tabular} &
  \begin{tabular}[c]{@{}c@{}}1.78 \\ (34.3\%)\end{tabular} \\ \hline
\textbf{Alexnet} &
  \begin{tabular}[c]{@{}c@{}}2.99 \\ (92.0\%)\end{tabular} &
  \begin{tabular}[c]{@{}c@{}}2.39 \\ (87.3\%)\end{tabular} &
  \begin{tabular}[c]{@{}c@{}}1.86 \\ (82.2\%)\end{tabular} &
  \multicolumn{1}{c|}{\begin{tabular}[c]{@{}c@{}}1.75 \\ (79.8\%)\end{tabular}} &
  \begin{tabular}[c]{@{}c@{}}2.99 \\ (89.6\%)\end{tabular} &
  \begin{tabular}[c]{@{}c@{}}2.39 \\ (71.5\%)\end{tabular} &
  \begin{tabular}[c]{@{}c@{}}1.86 \\ (55.8\%)\end{tabular} &
  \begin{tabular}[c]{@{}c@{}}1.75 \\ (56.3\%)\end{tabular} \\ \hline
\textbf{InceptionV3} &
  \begin{tabular}[c]{@{}c@{}}3.25 \\ (92.3\%)\end{tabular} &
  \begin{tabular}[c]{@{}c@{}}2.49 \\ (84.6\%)\end{tabular} &
  \begin{tabular}[c]{@{}c@{}}2.08 \\ (78.9\%)\end{tabular} &
  \multicolumn{1}{c|}{\begin{tabular}[c]{@{}c@{}}1.76 \\ (76.1\%)\end{tabular}} &
  \begin{tabular}[c]{@{}c@{}}3.25 \\ (82.1\%)\end{tabular} &
  \begin{tabular}[c]{@{}c@{}}2.49 \\ (75.8\%)\end{tabular} &
  \begin{tabular}[c]{@{}c@{}}2.08 \\ (50.0\%)\end{tabular} &
  \begin{tabular}[c]{@{}c@{}}1.76 \\ (39.1\%)\end{tabular} \\ \hline
\textbf{Resnet50} &
  \begin{tabular}[c]{@{}c@{}}3.02 \\ (92.0\%)\end{tabular} &
  \begin{tabular}[c]{@{}c@{}}2.33 \\ (86.1\%)\end{tabular} &
  \begin{tabular}[c]{@{}c@{}}2.02 \\ (81.6\%)\end{tabular} &
  \multicolumn{1}{c|}{\begin{tabular}[c]{@{}c@{}}1.75 \\ (77.9\%)\end{tabular}} &
  \begin{tabular}[c]{@{}c@{}}3.02 \\ (88.6\%)\end{tabular} &
  \begin{tabular}[c]{@{}c@{}}2.33 \\ (68.2\%)\end{tabular} &
  \begin{tabular}[c]{@{}c@{}}2.02 \\ (55.3\%)\end{tabular} &
  \begin{tabular}[c]{@{}c@{}}1.75 \\ (48.7\%)\end{tabular} \\ \hline
\end{tabular}
\end{table*}

\begin{table*}[]
\centering
\caption{QR result}
\begin{tabular}{ccccccccc}
\textbf{Error rate} &
  \textbf{5\%} &
  \textbf{10\%} &
  \textbf{15\%} &
  \textbf{20\%} &
  \textbf{5\%} &
  \textbf{10\%} &
  \textbf{15\%} &
  \textbf{20\%} \\ \hline
\textbf{\begin{tabular}[c]{@{}c@{}}Target \\ values\end{tabular}} &
  \multicolumn{4}{c|}{$Y_{all}$} &
  \multicolumn{4}{c}{$Y_{target}\geq3$} \\ \hline
\textbf{Mobilenet} &
  \begin{tabular}[c]{@{}c@{}}2.34 \\ (85.4\%)\end{tabular} &
  \begin{tabular}[c]{@{}c@{}}2.08 \\ (82.6\%)\end{tabular} &
  \begin{tabular}[c]{@{}c@{}}1.83 \\ (75.9\%)\end{tabular} &
  \multicolumn{1}{c|}{\begin{tabular}[c]{@{}c@{}}1.36 \\ (67.1\%)\end{tabular}} &
  \begin{tabular}[c]{@{}c@{}}2.21 \\ (56.3\%)\end{tabular} &
  \begin{tabular}[c]{@{}c@{}}1.98 \\ (46.0\%)\end{tabular} &
  \begin{tabular}[c]{@{}c@{}}1.76 \\ (31.3\%)\end{tabular} &
  \begin{tabular}[c]{@{}c@{}}1.15 \\ (26.5\%)\end{tabular} \\ \hline
\textbf{Alexnet} &
  \begin{tabular}[c]{@{}c@{}}2.67 \\ (92.7\%)\end{tabular} &
  \begin{tabular}[c]{@{}c@{}}2.38 \\ (85.8\%)\end{tabular} &
  \begin{tabular}[c]{@{}c@{}}1.65 \\ (78.6\%)\end{tabular} &
  \multicolumn{1}{c|}{\begin{tabular}[c]{@{}c@{}}1.50 \\ (76.2\%)\end{tabular}} &
  \begin{tabular}[c]{@{}c@{}}2.50 \\ (71.2\%)\end{tabular} &
  \begin{tabular}[c]{@{}c@{}}2.21 \\ (38.6\%)\end{tabular} &
  \begin{tabular}[c]{@{}c@{}}1.46 \\ (43.9\%)\end{tabular} &
  \begin{tabular}[c]{@{}c@{}}1.42 \\ (39.4\%)\end{tabular} \\ \hline
\textbf{InceptionV3} &
  \begin{tabular}[c]{@{}c@{}}2.37 \\ (77.9\%)\end{tabular} &
  \begin{tabular}[c]{@{}c@{}}2.40 \\ (74.9\%)\end{tabular} &
  \begin{tabular}[c]{@{}c@{}}1.86 \\ (71.1\%)\end{tabular} &
  \multicolumn{1}{c|}{\begin{tabular}[c]{@{}c@{}}1.48 \\ (63.2\%)\end{tabular}} &
  \begin{tabular}[c]{@{}c@{}}2.60 \\ (29.5\%)\end{tabular} &
  \begin{tabular}[c]{@{}c@{}}2.61 \\ (35.9\%)\end{tabular} &
  \begin{tabular}[c]{@{}c@{}}2.00 \\ (25.3\%)\end{tabular} &
  \begin{tabular}[c]{@{}c@{}}1.57 \\ (15.4\%)\end{tabular} \\ \hline
\textbf{Resnet50} &
  \begin{tabular}[c]{@{}c@{}}2.91 \\ (86.2\%)\end{tabular} &
  \begin{tabular}[c]{@{}c@{}}2.40 \\ (79.6\%)\end{tabular} &
  \begin{tabular}[c]{@{}c@{}}2.01 \\ (73.6\%)\end{tabular} &
  \multicolumn{1}{c|}{\begin{tabular}[c]{@{}c@{}}1.38 \\ (66.7\%)\end{tabular}} &
  \begin{tabular}[c]{@{}c@{}}3.39 \\ (55.8\%)\end{tabular} &
  \begin{tabular}[c]{@{}c@{}}2.68 \\ (40.4\%)\end{tabular} &
  \begin{tabular}[c]{@{}c@{}}2.15 \\ (31.1\%)\end{tabular} &
  \begin{tabular}[c]{@{}c@{}}1.26 \\ (19.7\%)\end{tabular} \\ \hline
\end{tabular}
\end{table*}

\begin{table*}[ht!]\label{CQR-6h}
\centering
\caption{CQR result}
\begin{tabular}{ccccccccc}
\textbf{Error rate} &
  \textbf{5\%} &
  \textbf{10\%} &
  \textbf{15\%} &
  \textbf{20\%} &
  \textbf{5\%} &
  \textbf{10\%} &
  \textbf{15\%} &
  \textbf{20\%} \\ \hline
\textbf{\begin{tabular}[c]{@{}c@{}}Target \\ values\end{tabular}} &
  \multicolumn{4}{c|}{$Y_{all}$} &
  \multicolumn{4}{c}{$Y_{target}\geq3$} \\ \hline
\textbf{Mobilenet} &
  \begin{tabular}[c]{@{}c@{}}2.70 \\ (89.9\%)\end{tabular} &
  \begin{tabular}[c]{@{}c@{}}2.19 \\ (84.6\%)\end{tabular} &
  \begin{tabular}[c]{@{}c@{}}1.87 \\ (77.2\%)\end{tabular} &
  \multicolumn{1}{c|}{\begin{tabular}[c]{@{}c@{}}1.60 \\ (74.3\%)\end{tabular}} &
  \begin{tabular}[c]{@{}c@{}}2.57 \\ (66.9\%)\end{tabular} &
  \begin{tabular}[c]{@{}c@{}}2.09 \\ (50.5\%)\end{tabular} &
  \begin{tabular}[c]{@{}c@{}}1.81 \\ (31.6\%)\end{tabular} &
  \begin{tabular}[c]{@{}c@{}}1.38 \\ (32.8\%)\end{tabular} \\ \hline
\textbf{Alexnet} &
  \begin{tabular}[c]{@{}c@{}}2.67 \\ (92.6\%)\end{tabular} &
  \begin{tabular}[c]{@{}c@{}}2.43 \\ (86.4\%)\end{tabular} &
  \begin{tabular}[c]{@{}c@{}}1.70 \\ (80.0\%)\end{tabular} &
  \multicolumn{1}{c|}{\begin{tabular}[c]{@{}c@{}}1.51 \\ (76.5\%)\end{tabular}} &
  \begin{tabular}[c]{@{}c@{}}2.50 \\ (71.2\%)\end{tabular} &
  \begin{tabular}[c]{@{}c@{}}2.27 \\ (41.2\%)\end{tabular} &
  \begin{tabular}[c]{@{}c@{}}1.51 \\ (46.2\%)\end{tabular} &
  \begin{tabular}[c]{@{}c@{}}1.43 \\ (39.6\%)\end{tabular} \\ \hline
\textbf{InceptionV3} &
  \begin{tabular}[c]{@{}c@{}}3.57 \\ (93.9\%)\end{tabular} &
  \begin{tabular}[c]{@{}c@{}}2.96 \\ (87.5\%)\end{tabular} &
  \begin{tabular}[c]{@{}c@{}}2.29 \\ (80.3\%)\end{tabular} &
  \multicolumn{1}{c|}{\begin{tabular}[c]{@{}c@{}}2.02 \\ (75.2\%)\end{tabular}} &
  \begin{tabular}[c]{@{}c@{}}3.81 \\ (69.7\%)\end{tabular} &
  \begin{tabular}[c]{@{}c@{}}3.16 \\ (53.0\%)\end{tabular} &
  \begin{tabular}[c]{@{}c@{}}2.43 \\ (38.4\%)\end{tabular} &
  \begin{tabular}[c]{@{}c@{}}2.11 \\ (29.3\%)\end{tabular} \\ \hline
\textbf{Resnet50} &
  \begin{tabular}[c]{@{}c@{}}3.11 \\ (91.6\%)\end{tabular} &
  \begin{tabular}[c]{@{}c@{}}2.73 \\ (86.1\%)\end{tabular} &
  \begin{tabular}[c]{@{}c@{}}2.23 \\ (79.0\%)\end{tabular} &
  \multicolumn{1}{c|}{\begin{tabular}[c]{@{}c@{}}1.69 \\ (76.8\%)\end{tabular}} &
  \begin{tabular}[c]{@{}c@{}}3.58 \\ (63.4\%)\end{tabular} &
  \begin{tabular}[c]{@{}c@{}}3.02 \\ (53.3\%)\end{tabular} &
  \begin{tabular}[c]{@{}c@{}}2.38 \\ (39.6\%)\end{tabular} &
  \begin{tabular}[c]{@{}c@{}}1.57 \\ (33.1\%)\end{tabular} \\ \hline
\end{tabular}
\end{table*}

\end{document}